\documentclass{iaus}
\usepackage{graphicx}
\usepackage{natbib}

\title[Interacting Galaxies in the Infrared] 
{Infrared Properties of Nearby Interacting Galaxies: from Spirals to
  ULIRGs}

\author[V. Charmandaris]   
{Vassilis Charmandaris$^1$%
  \footnote{Chercheur Associ\'e, Observatoire de Paris, F-75014,  Paris,
    France, and IESL/Foundation for Research and 
Technology-Hellas, PO Box 1527, 71110, Heraklion, Greece}}

\affiliation{$^1$Department of Physics, University of Crete, P.O. Box
  2208, Heraklion, 71003, Greece \break email: vassilis@physics.uoc.gr}

\pubyear{2006}
\volume{235}  
\pagerange{1--4}
\date{?? and in revised form ??}
\jname{Galaxy Evolution across the Hubble Time}
\editors{J. Palous, F. Combes, eds.}
\begin{document}

\maketitle

\begin{abstract}
I present a brief review of some of the mid-infrared properties of
interacting galaxies as these were revealed using observations from
the Infrared Space Observatory and Spitzer Space Telescope over the
last decade. The variation of the infrared spectral energy distribution
in interacting galaxies can be used as an extinction free tracer not
only of the location of the star formation activity but also of the
physical mechanism dominating their energy production.\footnote[2]{The
presentation is also available at {\it
http://www.physics.uoc.gr/$\sim$vassilis/vc\_iau235.pdf}}.
\end{abstract}

\firstsection 

\section{Introduction}

One of the major steps in the understanding of galaxy evolution was
the realization that tails and bridges are the result of galaxy
interactions \citep{Toomre72}. It was also proposed that the
morphology of the observed tidal features and the separation between
the galaxies could be used to create a ``merging sequence'' of (11 at
first) peculiar NGC galaxies found in the Arp Atlas. Ever since,
improvements in numerical modeling of the stellar and gaseous
component in galaxies have clearly demonstrated that galaxy
interactions cause large scale instabilities in the galactic disks
leading to the formation of transient bars which drive the gas into
the center of the galaxies \citep{Mihos96}.  Furthermore, numerous
multi-wavelength studies of those systems \citep[see][ and references
therein]{Hibbard96,Schweizer98} have been performed in effort to
better understand phenomena such as starburst and AGN activity, as
well as mass transfers and morphological transformations associated
with interacting galaxies. The focus of some of these studies was in
the identification of observational characteristics which could be
used as alternatives of assigning an ``age'' to the event of the
interaction \citep[i.e.][]{Schweizer92}. However, since it was
realized early on that the sites of most intense star formation were
typically obscured by dust
\citep[i.e.][]{Sanders88,Mirabel98,Charmandaris04}, the observed dust
emission was used not only as a proxy of star formation rate
\citep[see][]{Forster04}, but also the estimated temperature of the
various dust components was suggested as a tracer of the age of the
last major star forming event
\citep[i.e.][]{Charmandaris00,Georgakakis00,Xilouris04}. This is of
particular interest since there is ample evidence that the galaxy
interaction/merge rate increases sharply with redshift and that nearly
all of the Utraluminous Infrared Galaxies (ULIRGs), who as we now know
contribute substantially towards the energy production in the Universe
at z$>1.5$ \citep{Elbaz03}, are in fact interacting systems. The
launch of the Spitzer Space Telescope introduces major improvements in
sensitivity, wavelength coverage, and spatial resolution in
mid-infrared imaging and spectroscopy, and opens new avenues in using
the dust properties towards better understanding of the properties of
interacting galaxies.

\section{Results}

\subsection{ISO}

In order to quantify the variation of the global mid-infrared colors of
interacting galaxies --- beyond the seminal work of \citet{Bushouse88}
which was based on IRAS data --- a small sample of well known
interacting galaxies, for which the stage of interaction was known
{\it a priori}, was imaged by ISO. Extensive studies of the ISO/CAM
mid-infrared broad band colors indicated that one could use the ratio
of the LW3(12-18\,$\mu m$) to LW2(5-8.5\,$\mu$m) flux densities as an
indicator of the fraction of the very small grain continuum emission
continuum to the PAH feature emission. The ratio is close to unity for
quiescent star formation but it increases to values of 3 or more in
starburst systems. As we clearly see from Figure 1, this ratio is
correlated to the IRAS colors and varies monotonically with the
intensity of the star formation activity in those galaxies. This
result suggests that even though the bolometric luminosity of
interacting luminous infrared galaxies is found at $\lambda \geq 40
\mu m$, the study of the mid-infrared part of the spectrum can be used
to trace the location of the far-infrared peak.

\begin{figure}[h]
\resizebox{\hsize}{!}{\includegraphics{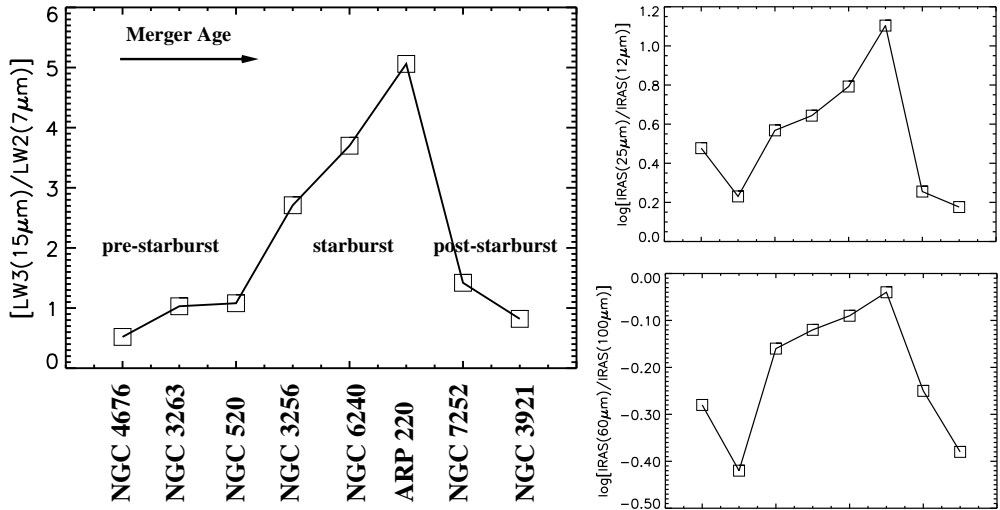}}
\caption{A comparison of the variation of the ISO/CAM 15$\mu$m/7$\mu$m
flux ratio along the merging sequence, with the well known IRAS flux
ratios. Note how well the ISO/CAM starburst diagnostic follows the
evolution of the star forming activity/merger age of the sequence. One
may effectively consider the 15$\mu$m/7$\mu$m ratio as a tracer of the
location of FIR peak of the bolometric luminosity
\citep[see][]{Charmandaris00}.}
\end{figure}

\subsection{Spitzer}

The first detailed study of the mid-infrared emission from a sample of
35 interacting galaxies with the large format detector arrays of
Spitzer, is presented by \citep{Smith06}. Some of those results are
highlighted in this section, while a number of other major imaging and
spectroscopy programs by Spitzer are underway. The \citep{Smith06}
sample was constructed from the Arp Atlas identifying nearby
($v<$11,000 km\,s$^{-1}$) binary systems with clear signs of tidal
disturbance and a total angular size of less than 3$'$ where each
component was greater than 30$''$. A control sample of ``normal''
isolated nearby galaxies obtained by the SINGS legacy program was used
for comparison \citep{Kennicutt03,Dale05}.

As indicated in Figure 2, the spatial resolution and sensitivity of
Spitzer enables us to quantify the distribution of the infrared
emission. We find that the 3.6$\mu$m emission from the tidal tails,
which can be used as a proxy of their stellar mass, is $\sim$7\% of
the galaxy disks. Furthermore, less than 10\% of the 24$\mu$m
emission, which identifies enshrouded regions of massive star
formation, is found outside the body of the interacting components.

\begin{figure}[h]
\resizebox{\hsize}{!}{\includegraphics{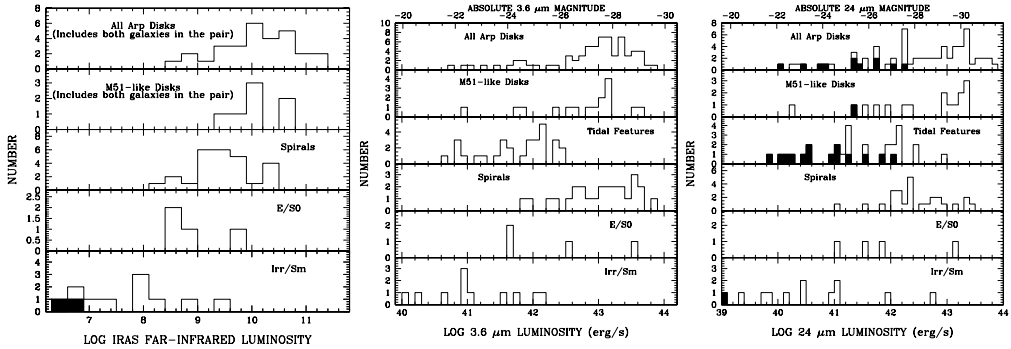}}
\caption{Histogram of the FIR luminosity, in log[L$_{\odot}$], of the
  interacting galaxy sample based on the global IRAS data, as well as
  the 3.6$\mu$m and 24$\mu$m based on the Spitzer data
  \citep[see][]{Smith06}. The solid areas indicate upper limits. Note
  the extent of the 3.6 and 24$\mu$m luminosities of the tidal
  features.}
\end{figure}

\begin{figure}[!ht]
\resizebox{\hsize}{!}{\includegraphics{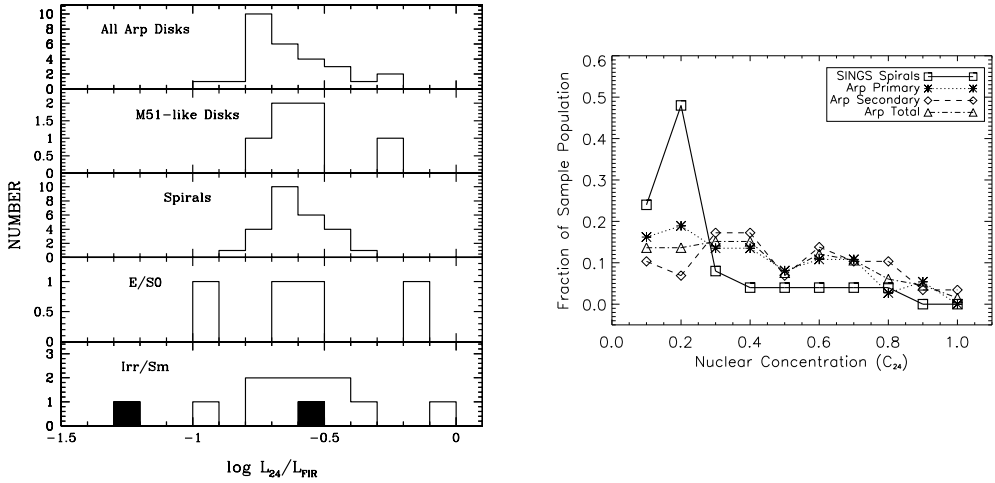}}
\caption{{\it Left:} Histogram of the ratio of the 24$\mu$m to FIR
  luminosity for all galaxies in the sample. {\it Right:} The 24$\mu$m
  central concentration index \citep[see][]{Smith06}}
\end{figure}

Analyzing the mid-infrared colors and luminosities of the interacting
and control sample, \citet{Smith06} showed that the star formation rate
(SFR) in the interacting galaxy sample was relatively modest, $\sim$ 1
M$_{\odot}$yr$^{-1}$, similar to what was found for the isolated
``normal'' galaxies. There was no apparent variation in the
distribution of mid-infrared colors for the various regions suggesting
that the normalized SFR in tails and bridges is similar to that found
on galaxy disks. No variation was either found as a function of the
separation between the two interacting components. It is likely though
that this result is simply due to the fact that systems in late stages
of interactions/merging were excluded from the interacting sample.

Despite these similarities between the two sample, in order to
investigate whether the Arp pairs have more centrally concentrated
star formation, \citet{Smith06} evaluated the 24$\mu$m nuclear
emission in a region of diameter 2~kpc centered on the position of the
nucleus of each galaxy and defined the nuclear concentration C$_{24}$
to be the ratio of the flux in the aperture S$_{\rm 2kpc}$ to the
total flux S$_{tot}$ from the galaxy. As shown in Figure 3 there is a
clear difference in the distribution of 24$\mu$m nuclear
concentrations between the spirals and the Arp systems. The spirals
are predominantly diffuse objects with the median value of C$_{24}$
being around 0.2 (i.e., 20$\%$ of the flux coming from the inner 2kpc)
and only tiny fraction of the systems having nuclear concentrations,
whereas the Arp systems show an almost flat (and slowly declining)
distribution of C$_{24}$. In particular, large numbers of Arp systems
(both primary and secondary galaxies) have values of C$_{24}$ greater
than 0.3 and in some cases are totally nuclear dominated. This
suggests that even though the Arp systems have similar L$_{24\mu
m}$/L$_{\rm FIR}$ luminosities they show a much wider distribution of
24$\mu$m nuclear concentrations, with a large excess (over the
spirals) extending up to objects which are essentially point sources.

\vspace*{-0.3cm}
\section{Conclusions}
The large sample of galaxies that can now be efficiently studied in
the mid-infrared with Spitzer is enabling us to establish for the
first time well calibrated extinction free star formation diagnostics
in galaxies.

\begin{acknowledgments}
I would like to acknowledge the ISO/CAM extragalactic team at
CEA/Saclay and the Spitzer/IRS instrument team at Cornell and Caltech
for stimulating discussions over the past decade.
\end{acknowledgments}


\end{document}